\documentclass[11pt,a4paper]{article}
\usepackage[final]{amlc2019}
\usepackage{booktabs}       % professional-quality tables
\usepackage{nicefrac}       % compact symbols for 1/2, etc.
\usepackage{times}
\usepackage{latexsym}
\usepackage{url}
\usepackage{amsmath}
\usepackage{graphicx}
\usepackage{hyperref}  
\usepackage[bottom]{footmisc}

\title{Knowledge Distillation in Document Retrieval}

\author{%
  Siamak Shakeri \thanks{corresponding author}\\
  Amazon Alexa AI\\
  \texttt{siamaks@amazon.com} \\
  \And
   Abhinav Sethy \\
  Amazon Alexa AI\\
  \texttt{sethya@amazon.com} \\
  \And
    Cheng Cheng \thanks{work done while interning at Amazon Alexa AI} \\
  Facebook\\
  \texttt{cc2@illinois.edu} \\
}

\date{}

\begin{document}
\maketitle
\begin{abstract}
Complex deep learning models now achieve  state of the art performance for many document retrieval tasks. The best models process the query or claim jointly with the document. However for fast scalable search it is desirable to have document embeddings which are independent of the claim. In this paper we show that knowledge distillation can be used  to encourage a  model that generates claim independent document encodings to mimic the behavior of a more complex model which generates  claim dependent encodings. We explore this approach in document retrieval for a fact extraction and verification task. We  show that by using the soft labels from a complex cross attention  teacher model, the performance of claim independent student LSTM or CNN models is improved across all the ranking metrics. The student models we use are 12x faster in runtime and 20x smaller in number of parameters than the teacher.
\end{abstract}

\section{Introduction}\label{intro}

Deep learning models have shown promising results in the field of document retrieval. Specifically attention based models such as (\cite{attentionattention, bert, DCN}) demonstrate clear improvements in the performance of neural models in question answering tasks. In such models, rich encodings of claim (question) and document (answers) are generated using various attention mechanisms. A challenge when using such models in large scale document retrieval systems is the lack of separation between document and claim encodings, making it infeasible to pre-index and retrieve the   document encodings efficiently during runtime. In this paper we explore the use of knowledge distillation as a means to transfer the embedded attention information to a simpler attention-free neural model.

Knowledge distillation using posterior probabilities of one model to improve the performance of another model has been widely studied (\cite{bucila}). (\cite{hinton}) discusses using aggregate posteriors of an ensemble of acoustic deep models to improve the performance of a single model.
(\cite{yoonkim}) suggests using word-level knowledge distillation in Neural Machine Translation.
(\cite{zagoruyko}) defines an attention mechanism in Convolutional Neural Networks and uses knowledge distillation to improve the performance of a student model by forcing it to mimic such mechanism. 
(\cite{fitnet}) explores training a student model which is deeper and thinner than the teacher while utilizing both softmax posteriors and intermediate layer representations of the teacher.
(\cite{distillwv}) has experimented with distilling knowledge from a large embedding to a smaller one.
(\cite{attentionguided}) have used an ensemble of models as the teacher model, similar to (\cite{hinton}), to guide the alignments of the student model in machine reading comprehension.

 We conduct knowledge distillation experiments on document retrieval for the fact extraction and verification task introduced  in (\cite{Thorne18Fever}). In order to make our approach generic, no restrictions are imposed on the type of attention that the teacher model can employ. Furthermore, the student model does not need to be the same type as the teacher model, e.g. the teacher model can be a CNN based model while the student is an LSTM. The student models that are experimented with in this paper are both faster (up to 12x) and smaller (up to 20x) than the teacher model. We start with the problem definition and task setup in section~\ref{prob_descript}. Next we look at model training with knowledge distillation in section~\ref{sec:training} and present experimental results in section~\ref{sec:results}.

\section{Problem Description}\label{prob_descript}

\begin{figure}[!htb]
        \center{\includegraphics[width=1\textwidth]
        {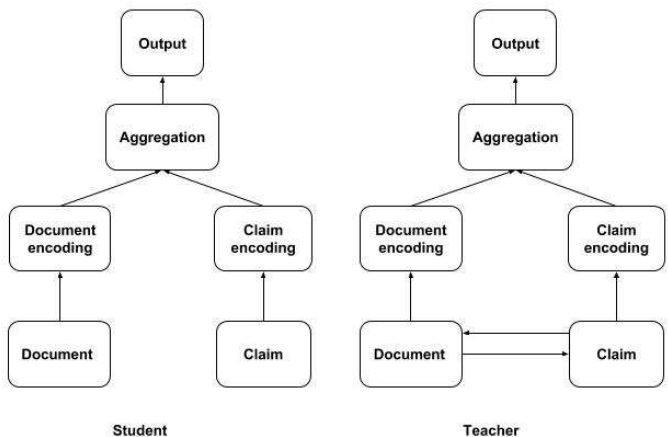}}
        \caption{Teacher vs Student Model}
        \label{fig:teacher_student}
      \end{figure}

In the knowledge distillation or teacher-student training framework, the \textit{student model} is the target model to be trained using annotation labels and information such as posteriors or hidden unit activations from a complex teacher model. In this paper we consider single layer CNN and LSTM models with a linear layer on top as our student models. Specifically, we avoid models that require interactions between claim and document to create encodings of both the document and claim. \textit{Teacher model} is a more complex model than the student model both in terms of the number of parameters and the structure of the network. As shown in \ref{fig:teacher_student}, the teacher model uses claim dependent document encodings.

The document retrieval task can be considered as a classification task: Given a $< document, claim>$ pair, shortened as $<d, c>$, assign a score indicating the relevancy of the document to the claim. For each claim, the documents are sorted based on the assigned score, and the top ones are picked. We further discuss the metrics used in later sections.
\subsection{Data}
The publicly available FEVER dataset is used in this paper (\cite{Thorne18Fever}). In FEVER, a corpus of Wikipedia documents is given, and the task is to classify a given claim as \textit{supported}, \textit{refuted} or \textit{not enough info} using the given corpus. Three sub tasks are defined: document retrieval, sentence retrieval and textual entailment. In this paper, we focus on the document retrieval task. The corpus consists of 5.4 million pages, and more than 175,000 claims. Each sample in FEVER consists of a claim, all the relevant documents, all the relevant sentences in those documents, and the annotated label.

For training and evaluating our model, we construct $<d, c, label>$ tuples. For each claim, all the annotated relevant documents are labeled as positive samples. DrQA (\cite{drqa}) is employed to find the $k$ nearest documents of a claim from the entire corpus based on cosine similarity of TF-IDF vectors. The top results returned by DrQA that are not annotated as relevant are labeled as negative samples. The rationale behind this is to have most similar irrelevant documents to the claim as negative samples. This makes the resulting dataset to be non-trivial. Each claim has a fixed number of documents \textit{C}. The claims are split into train, dev and test sets, each having 145000, 20000 and 10000 claims respectively.

Table \ref{tab:k} shows given a certain \textit{C}, what percentage of claims will have all the annotated relevant documents. We use \textit{C}=10, as it will cover vast majority of the claims.
\begin{table}
\centering
%\small
\begin{tabular}{|c| c | c | c| c | c | } 
\hline
\textit{C} & 1 & 2 & 5 & 10 & 15 \\ 
\hline
claims & 87.37 & 96.90 & 99.27 & 99.90 & 99.99 \\ 
\hline
\end{tabular}
\caption{\footnotesize{Percentage of claims with all the relevant documents versus \textit{C}}.}
\label{tab:k}
\end{table}
\subsection{Metrics}
Being a ranking task, Discounted Cumulative Gain (DCG) and Recall at top \textit{k} values are the performance metrics. In order to aggregate per-claim recall values, we define the followings:
\vspace{-2mm}
\begin{itemize}
    \item recall\textsubscript{macro}(k)=  $\frac{\sum_{i=1}^{N}\{\frac{\sum_{j=1}^{k}r_{ij}}{c_{i}}\}}{N} $
    \item recall\textsubscript{micro}(k)= $\frac{\sum_{i=1}^{N}\sum_{j=1}^{k}r_{ij}}{\sum_{i=1}^{N}c_{i}} $
\end{itemize}
\vspace{-2.5mm}

Where, $c_{i}$ indicates the number of relevant documents for claim $i$, $r_{ij}$ is $1$ if document at $j$th position in the sorted documents list is relevant to claim $i$  and $0$ otherwise. $N$ indicates the total number of claims. 

\section{Training}
In this section we discuss the training setup for our knowledge distillation experiments.
\label{sec:training}
\subsection{Teacher Model}
As our teacher models, we experimented with architectures that have been shown to give state of the art performance on the SQuAD task (\cite{squad}). Two models that performed the best were DCN (\cite{DCN}) without highway maxout network layers and Transformer (\cite{transformer}). Note that the purpose of our study is not to find the best teacher model, but a teacher model that significantly outperforms the baseline student model. Table \ref{tab:teacher} shows the performance of these two models. We picked the DCN model as the teacher for further experiments. Please note that the models were modified to be used in our $\textless d, c\textgreater$  classification task. 

\begin{table}
\centering
%\small
\begin{tabular}{|c|c|c| c | c | c| c |} 
\hline
 Model & R(1)& R\textsubscript{micro}(3) & R\textsubscript{macro}(3) & R\textsubscript{micro}(5) & R\textsubscript{macro}(5) & DCG \\ 
\hline
DCN & 63.79 & 82.21 & 84.83 & 92.12 & 93.61& 91.56 \\
\hline
Transformer & 43.29 & 72.25 & 75.49 & 89.47& 91.25& 89.72 \\ 
\hline
\end{tabular}
\caption{\footnotesize{Teacher Models Metrics}}
\label{tab:teacher}
\end{table}
\subsection{Student Model}\label{sec:student}
The following candidate models were employed as student:

\textit{SimpleCNN}: CNN and Maxpooling layers are separately applied to claim and document to create the encodings. A linear layer is used to join the encodings.

\textit{SimpleLSTM}: Recurrent layers separately applied to claim and document to create the encodings. Similar to SimpleCNN, A linear layer is used to join the encodings.

The final claim and document encodings are independent of each other, as mentioned in \ref{prob_descript}.

\subsection{Objective Function}
In order to train the student model, the trained teacher model is run over the entire training, dev and test sets, and similarity score of each $\textless d, c\textgreater$  pair is recorded. When training the student network, these similarity scores alongside the annotated labels are used. We define the following losses:

{
\begin{align}
    &t'_{oi} = \frac{e^{\frac{t_{oi}}{T}}}{\sum_{j=1}^{C}e^{\frac{t_{oj}}{T}}}\\
    & \textrm{SoftLoss}_{CE}: \: \sum_{i=1}^{C}t'_{oi}\log s_{oi} \quad 
    \textrm{SoftLoss}_{MSE}: \: \sum_{i=1}^{C}||t'_{oi} - s_{oi}||^{2} \quad 
    &\textrm{HardLoss}: \: \sum_{i=1}^{C}y_{oi}\log s_{oi} \\
    &\textrm{Loss}: \: \alpha \times \textrm{SoftLoss}_{(CE\: or \: MSE)} 
    + (1-\alpha)\times \textrm{HardLoss}
\end{align} 
}
Where, $t_{oj}, s_{oj}, y_{oj}$ denote teacher logits, student logits and true label of document $j$ and claim $o$, respectively. $\alpha$ is a hyper parameter that dictates the importance of $SoftLoss$. Setting it to 0 indicates no teacher training. $T$ (Temperature) is another hyper parameter that indicates how much smoothing of the classification scores is done. Setting it to 0 is equal to picking the largest value only.
\section{Experimental Results}
\subsection{Full Training Data}

\label{sec:results}
\begin{table}
\centering
%\small
\begin{tabular}{|c|c|c|c|c|c|c|} 
\hline
 SoftLoss/$\alpha$/T & R(1) & R\textsubscript{micro}(3) & R\textsubscript{macro}(3) & R\textsubscript{micro}(5) &R\textsubscript{macro}(5) & DCG\\ 
\hline
 \textbf{No Teacher}  & 42.27&  72.49 & 69.43 & 87.46 & 85.46 & 79.72\\
\hline
MSE/1.0/4  &44.88 &74.79 &71.66 &89.31 &87.71 & 81.38\\
\hline
\textbf{MSE/0.2/3}  &46.16 &\textbf{76.77} &\textbf{73.69} &\textbf{90.23} &\textbf{88.42} &\textbf{82.13} \\
\hline
MSE/0.65/4 &44.15 &74.28 &71.26 &89.35 &87.74 &80.86 \\
\hline
CE/0.2/3 &43.57 &75.89 &72.89 &89.42 &87.56 &80.93 \\
\hline
CE/0.5/6 &\textbf{46.91} &74.36 &71.22 &88.57 &86.66 &81.87 \\
\hline
CE/0.4/5 &40.71 &72.75 &69.49 &88.66 &86.53 &78.87 \\
\hline
\end{tabular}
\caption{\footnotesize{Teacher Student Training with SimpleLSTM. \textit{Softloss/$\alpha$/T} indicates the softloss type used, SoftLoss importance and temperature, respectively.}}
\label{tab:simpleLSTM}
\end{table}
\begin{table}
\centering
%\small
\begin{tabular}{|c|c|c|c|c|c|c|} 
\hline
 SoftLoss/$\alpha$/T & R(1) & R\textsubscript{micro}(3) & R\textsubscript{macro}(3) & R\textsubscript{micro}(5) &R\textsubscript{macro}(5) & DCG\\ 
\hline
\textbf{No Teacher}  &\textbf{44.13} & 70.17 & 65.98 & 84.28 & 81.2  & \textbf{83.95}\\
\hline
MSE/0.8/6  &42.95 &70.3 & 66.4 & 84.84 & \textbf{82.39} & 83.52\\
\hline
MSE/1.0/1 &41.07 &68.77 & 64.4 & 84.63 & 81.74 & 82.26 \\
\hline
MSE/0.8/2 &40.31 & 67.89 &63.5 &84.62 &81.55 &81.81 \\
\hline
\textbf{CE/0.5/3} & 43.59 & \textbf{71.16} & \textbf{67.13} & \textbf{85.11} & 82.28 & 83.94\\
\hline
CE/0.65/1 & 41.96 & 67.18 & 63.07 & 83.29 & 80.2 & 82.4 \\
\hline
CE/0.3/2 & 41.0 & 66.67 & 62.51 & 82.14 & 79.01 & 81.65 \\
\hline
\end{tabular}
\caption{\footnotesize{Teacher Student Training with SimpleCNN}}
\label{tab:simpleCNN}
\end{table}
\begin{table}
\centering
%\footnotesize
\begin{tabular}{|c|c|c|c|c|c|c|c|} 
\hline
  & \textsubscript{Model}& R(1) & R\textsubscript{micro}(3) & R\textsubscript{macro}(3) & R\textsubscript{micro}(5) &R\textsubscript{macro}(5) & DCG\\ 
\hline
 {Best Diff} &{\textit{A}} &3.89 &4.28 &4.26 &2.77 &2.96 & 2.41\\
\hline
{MSE Avg} &{\textit{A}} & 45.06& 75.28& 72.20 & 89.63 & 87.96& 81.46\\
\hline
{CE Avg} &{\textit{A}} &43.73 &74.33 &71.2 &88.88 &86.92 &81.22 \\
\hline
{Best Diff} &{\textit{B}} &-0.54 & 0.99 & 1.15 & 0.83 & 1.08 & -0.01\\
\hline
{MSE Avg} &{\textit{B}} & 41.44 & 68.99 & 64.77 & 84.7 & 81.89 & 82.53\\
\hline
{CE Avg} &{\textit{B}}& 42.18 & 68.34 & 64.24 & 83.51 & 80.50 & 82.66\\
\hline
\end{tabular}
\caption{\footnotesize{Teacher Student Training Averages. \textit{Best Diff} indicates the difference between the best student model when teacher training is used vs when training with only hard labels. \textit{MSE} and \textit{CE Avg}s indicate average across the top three performing configurations when using MSE and CE softlosses, respectively. \textit{A} refers to SimpleLSTM and \textit{B} refers to SimpleCNN}}.
\label{tab:simpleLSTM}
\end{table}

\begin{table}
\centering
%\small
\begin{tabular}{|c|c|c|c|c|c|c|} 
\hline
 SoftLoss/$\alpha$/T & R(1) & R\textsubscript{micro}(3) & R\textsubscript{macro}(3) & R\textsubscript{micro}(5) &R\textsubscript{macro}(5) & DCG\\ 
\hline
No Teacher  &39.69 & 69.95 & 66.56 & 85.57 & 83.21  & 77.6\\
\hline
CE/1.0/3  &40.83 &69.35 & 65.74 & 85.42 & 83.02 & 77.65\\
\hline
MSE/0.65/5 & 37.82 & 68.45 & 64.85 & 85.21 & 82.54 & 76.39 \\
\hline
\end{tabular}
\caption{\footnotesize{Teacher Student Training with SimpleLSTM, $10\%$ Data}}
\label{tab:simpleLSTM10}
\end{table}

\begin{table}
\centering
%\small
\begin{tabular}{|c|c|c|c|c|c|c|} 
\hline
 SoftLoss/$\alpha$/T & R(1) & R\textsubscript{micro}(3) & R\textsubscript{macro}(3) & R\textsubscript{micro}(5) &R\textsubscript{macro}(5) & DCG\\ 
\hline
No Teacher  & 33.11 & 62.89 & 59.28 & 81.74 & 78.97  & 73.03\\
\hline
CE/1.0/2  &32.66 & 64.57 & 61.16 & 83.7 & 80.96 & 73.23\\
\hline
MSE/0.3/6 &33.29 &64.64 & 61.28 & 82.08 & 79.36 & 73.25 \\
\hline
\end{tabular}
\caption{\footnotesize{Teacher Student Training with SimpleLSTM, $3\%$ Data}}
\label{tab:simpleLSTM3}
\end{table}
We first experiment with training the teacher model (\ref{sec:training}) with the entire training data, and then using the posteriors in training the student models (\ref{sec:student}). Tables \ref{tab:simpleCNN} and \ref{tab:simpleLSTM}   show the top performing results. For each loss type and model, the top three performing models are picked. Some observations are as follows:
\begin{itemize}
  \setlength\itemsep{0em}
    \item Using teacher student training improves the performance of student models.
    \item Improvements resulting from knowledge distillation are larger with SimpleLSTM. This indicates that the LSTM module is more capable of benefiting from the information embedded in the soft labels provided by the teacher model, as well as its superiority in encoding sequential inputs (\cite{lstm}), (\cite{bahdanau})
    \item Best performance is achieved with temperatures $>$ 1. This shows that using smoothing of the logits is crucial. (\cite{hinton}) also shows improvements using smoothing. In fact, none of top performing runs have been with $T < 1$.
    \item The best performance is achieved when using a mix of teacher and hard labels. It can be seen that $1 > \alpha > 0$ values generate the largest improvements. Using only soft labels from teacher model lacks the more credible annotated labels. Using only hard labels lacks the extra information provided by soft labels. This indicates soft and hard labels provide complementing information.
    \item \textit{MSE vs CE:} Results do not show any consistent pattern of distillation favoring one versus the other. For SimpleLSTM, MSE performs better, and for SimpleCNN, CE is a better choice.
 \end{itemize}
 
\subsection{Partial Training Data}
It has been claimed (\cite{hinton}) that teacher student training could act as a regularizer. We test this claim by designing an experiment where using a small portion of the entire dataset to train, we expect less overfitting when employing knowledge distillation versus when no teacher training is involved.

In this section, the results of experiments with only partial training data to train the student model are discussed. Please note that the teacher model is trained with the entire training dataset. We experimented with SimpleLSTM model training it with only $10\%$ and $3\%$ of the training set.

Tables \ref{tab:simpleLSTM10} and \ref{tab:simpleLSTM3} show the results. $<2\%$ improvements with $3\%$ training and none with $10\%$ training are observed. These results do not support the hypothesis regarding the teacher training being a regularizer. The results show by having more training data, the benefits of the soft labels would become more evident.
\subsection{Running time}
The experiments were done on AWS EC2 instances running on Tesla V100 GPUs. PyTorch (\cite{pytorch}) was employed to implement the neural models. The student model is up to 12x faster and 20x smaller in the number of parameters than the teacher. This is besides the reduction in computational complexity by reusing the indexed document encodings as discussed in \ref{intro}. Particularly, if there are \textit{D} documents and \textit{N} claims where each document should be evaluated for each of the claims, computation cost of student model is \textit{O(N+D)} while teacher's  is \textit{O(ND)}. Please note that the cost is in the unit of computing the encoding of claim or document. Table \ref{tab:runtime} shows detailed running time metrics.
\begin{table}
\centering
%\small
\begin{tabular}{|c|c|c|c|} 
\hline
Model& \# Parameters&Loading Time&Evaluation Time \\
\hline
Teacher&2.85M& 5.35s& 760.7s \\
\hline
SimpleLSTM&141k& 5.1s & 66.5s \\
\hline 
\end{tabular}
\caption{\footnotesize{Run time Performance with Batch Size of 8}}
\label{tab:runtime}
\end{table}

\section{Conclusion}
In this paper, we proposed using knowledge distillation to improve the performance of student models that generate claim independent document encodings in document retrieval task for factual verification. We experimented with various configurations when adding the teacher model posteriors to the student training, and results show that significant improvements can be achieved across the ranking metrics, without sacrificing runnig time advantages of simpler models.
In future, we propose applying this work to a larger set of input documents (\textit{C}) to replace the DrQA retriever with the student model.
\newpage
\newpage

\bibliographystyle{amlc_natbib}
\bibliography{amlc2019}

% \appendix

% \section{Appendices}
% \label{sec:appendix}

% \section{Supplemental Material}

\end{document}